\documentclass[reprint,prd,twocolumn,superscriptaddress,showpacs,nofootinbib,floatfix,preprintnumbers]{revtex4-1}
\usepackage{graphicx, bm, color}
\newcommand{\lsim}{\mbox{\raisebox{-.6ex}{~$\stackrel{<}{\sim}$~}}}
\newcommand{\gsim}{\mbox{\raisebox{-.6ex}{~$\stackrel{>}{\sim}$~}}}
\begin{document}
\preprint{MAN/HEP/2013/22}
\title{
Standard Model Explanation of the Ultra-high Energy Neutrino Events at IceCube}
\author{Chien-Yi Chen}
\affiliation{Department of Physics, Brookhaven National Laboratory, Upton, New York 11973, USA}

\author{P. S. Bhupal Dev}
\affiliation{Consortium for Fundamental Physics, School of Physics and Astronomy, University of Manchester, Manchester M13 9PL, United Kingdom}

\author{Amarjit Soni}
\affiliation{Department of Physics, Brookhaven National Laboratory, Upton, New York 11973, USA}
\begin{abstract}
The recent observation of two PeV events at IceCube, followed by an additional 26 events between 30 - 300 TeV, has generated considerable speculations on its origin, and many exotic New Physics explanations 
have been invoked. For a reliable interpretation, it is however important 
to first scrutinize the 
Standard Model (SM) expectations carefully, including the 
theoretical uncertainties, mainly due to the parton distribution functions. 
Assuming a new isotropic cosmic neutrino flux with a simple unbroken power-law spectrum, $\Phi\propto E^{-s}$ for the entire energy range of interest, we find that with $s=1.5$ - 2, the 
SM neutrino-nucleon interactions are sufficient to explain all the observed 
events so far, without the need for any beyond the SM explanation. 
With more statistics, this powerful detector could provide a unique test of the SM up to the PeV scale, and lead to important clues of New Physics.    
\end{abstract}
\maketitle
\section{Introduction} \label{sec:1}
Recently the IceCube collaboration has 
reported the first observation of extremely 
intriguing two events with neutrino energies above 1 
PeV~\cite{Aartsen:2013bka}.  These are by far 
the highest neutrino energies that have  ever been observed.  As such they 
could potentially represent the first detection of a non-atmospheric, 
astrophysical high energy neutrino flux,  thus opening an avenue for a number of astrophysical objects and mechanisms  to provide information complementary to that obtained from electromagnetic  or hadronic observations~\cite{Becker:2007sv}.  In a follow-up search with improved sensitivity and extended energy coverage, 
IceCube has reported 
additional 26 events with deposited energies ranging from 30 to 300 TeV~\cite{science}.  
These 28 events are the result of the first 662 days of data-taking, and give a preliminary significance of $4.1\sigma$ with respect to the reference atmospheric neutrino background model.  
These observations may well hold the key to 
understanding neutrino masses, the nature of neutrino mass-hierarchy, their 
relevance to Dark Matter, or in general, to physics beyond 
the Standard Model (SM).  For these  reasons, it is extremely important to 
understand every possible aspect of these IceCube events. 

This realization has spurred a lot of interest on the origin of these ultra-high energy (UHE) neutrino events and their underlying spectral shape. Various extra-terrestrial sources (e.g., gamma-ray bursts, active galactic nuclei, early supernovae, baby neutron stars, starburst galaxies, cosmogenic)~\cite{astro} with different power-law fluxes have been analyzed. From a particle physics point of view, several beyond SM phenomena, e.g., decaying heavy Dark Matter~\cite{Feldstein:2013kka}, lepto-quark resonance~\cite{Barger:2013pla}, decay of massive neutrinos to light ones over cosmological distances~\cite{Pakvasa:2012db}, and pseudo-Dirac neutrinos oscillating to their sterile counterparts in a mirror world~\cite{Joshipura:2013yba},  
have been proposed. Most of these proposals are motivated by some specific features in the IceCube 
data such as a (slight) paucity of muon tracks, a (possible) apparent 
energy gap between 300 TeV and 1 PeV, and possibly a bit higher than expected event rate above PeV. 

Our primary aim in this paper is to carefully scrutinize the IceCube 
observations with respect to the SM expectations, taking into account the theoretical uncertainties, mainly due to the parton distribution functions (PDFs). Following the IceCube analysis~\cite{science}, which did not find significant clustering of the events in time or space compared to randomized sky maps, 
we assume a simple isotropic astrophysical power-law spectrum for the UHE neutrino flux $\Phi\propto E^{-s}$,  
and show that for $s=1.5$ - 2, the current data is consistent with the SM, within the 
theoretical and 
experimental uncertainties. Thus there is no significant feature of the current 
data requiring an exotic particle physics explanation, other than of course a very interesting new cosmic neutrino flux. 
However, we want to stress that there are mild indications of slight potential 
deficit of muons in comparison to other flavors and perhaps a little excess of 
above PeV events. If these features attain clear statistical significance as 
more data is accumulated, then some new physics interactions 
may well become necessary; 
but at this juncture, these 
considerations appear somewhat premature.

The rest of the paper is organized as follows: In Section~\ref{sec:2}, we calculate 
various neutrino-induced scattering cross sections in the SM, along with their differential distributions, for a reference PDF set, and compare the predictions for the central value as well as the 90\% confidence level (CL) range of the PDFs 
at Leading Order (LO), Next-to-Leading Order (NLO) and Next-to-Next-to-Leading Order (NNLO). In Section~\ref{sec:3}, we use the IceCube experimental parameters to carefully estimate the corresponding number of events predicted by the SM, along with its theoretical uncertainty, and compare our results with the IceCube observed events for a simple power-law cosmic neutrino flux. We also perform a $\chi^2$-analysis to find 
the best-fit spectral index and normalization for this new flux. 
Finally, our conclusions are given in Section~\ref{sec:4}.  
\section{Standard Model Cross Section}\label{sec:2}
The expected neutrino-induced event rate at IceCube can be schematically written as 
\begin{eqnarray}
\frac{dN}{dE_{\rm EM}}=T\cdot \Omega \cdot N_{\rm eff}(E_\nu) \cdot \sigma(E_\nu) \cdot\Phi (E_\nu)
\label{eq:N}
\end{eqnarray}
where $E_\nu$ is the incoming neutrino energy in the {\it laboratory frame}, $E_{\rm EM}$ is the electromagnetic (EM)-equivalent deposited energy, $T$ is the time of exposure, $\Omega$ is the solid angle of coverage, $N_{\rm eff}$ is the effective number of target nucleons,  $\sigma$ is the neutrino-induced interaction cross section, and $\Phi$ is the incident neutrino flux. 

The main particle physics ingredient in Eq.~(\ref{eq:N}) is the neutrino-induced interaction cross section~\cite{Formaggio:2013kya}. Within the SM, neutrinos interact with matter only through the exchange of weak  gauge bosons. The dominant processes (in most of the energy range of interest) are the charged-current (CC) and neutral-current (NC) neutrino-nucleon deep inelastic scattering (DIS) mediated by $t$-channel $W$ and $Z$ respectively: 
\begin{eqnarray}
\nu_\ell+N &\to & \ell+X~~~~({\rm CC}), \label{cc1} \\
\nu_\ell+N & \to & \nu_\ell+X~~~~({\rm NC}), \label{nc1}
\end{eqnarray}
where $\ell=e,\mu,\tau$ represents the $SU(2)_L$ lepton-flavor, $N=(n+p)/2$ is an isoscalar nucleon ($n$ and $p$ being the neutron and proton, respectively) in the renormalization group-improved parton model, 
and $X$ is the hadronic final state. The neutrino interactions with the electrons in the target material can generally be neglected with respect to the neutrino-nucleon interactions due to the electron's small mass ($m_e\ll M_N$). There is however an important exception for the $\bar\nu_ee^-$ interaction when the incoming neutrino energy is between about 4 - 10 PeV. In this case, the resonant production of the SM $W$-boson gives a 
significant enhancement in the $\bar{\nu}_ee$ 
cross section, which peaks at $E_\nu^{\rm res}=M_W^2/2m_e=6.3$ PeV, and is commonly 
known as the {\it Glashow resonance}~\cite{glashow}. 
\subsection{Differential Cross Sections}\label{sec:2a}
The neutrino-nucleon differential scattering 
cross sections at leading order are given by~\cite{gandhi} 
\begin{eqnarray}
\frac{d^2\sigma_{\nu N}^{\rm CC}}{dxdy} &=& \frac{2G_F^2 M_N E_\nu}{\pi}
\left(\frac{M_W^2}{Q^2+M_W^2}\right)^2\nonumber\\
&&\times \left[xq(x,Q^2)+x\bar{q}(x,Q^2)(1-y)^2\right],\label{cc}\\
\frac{d^2\sigma_{\nu N}^{\rm NC}}{dxdy} &=& \frac{G_F^2 M_N E_\nu}{2\pi}
\left(\frac{M_Z^2}{Q^2+M_Z^2}\right)^2\nonumber\\
&& \times \left[xq^0(x,Q^2)+x\bar{q}^0(x,Q^2)(1-y)^2\right],\label{nc}
\end{eqnarray}
where $-Q^2$ is the invariant momentum-square transfer to the exchanged vector boson, $M_N$ and $M_{W(Z)}$ are the nucleon and intermediate $W(Z)$-boson masses respectively, and $G_F$ is the Fermi coupling constant. The differential distributions in Eqs.~(\ref{cc}) and (\ref{nc}) are with respect to the Bjorken scaling variable $x$ and the inelasticity parameter $y$, where 
\begin{eqnarray}
x=\frac{Q^2}{2M_N yE_\nu}~~~{\rm and}~~~y=\frac{E_\nu-E_\ell}{E_\nu}, \label{xy}
\end{eqnarray}
$E_\ell$ being the energy 
carried away by the outgoing lepton in the laboratory frame, and $x$ is the fraction of the initial nucleon momentum taken by the struck quark. 
Here $q,\bar{q}$ ($q^0,\bar{q}^0$) are respectively the quark and anti-quark 
density distributions  in a proton, summed over valence and sea quarks of all 
flavors relevant for CC (NC) interactions~\cite{gandhi}:  
\begin{eqnarray}
q &=& \frac{u+d}{2}+s+b,\\
\bar{q} &=& \frac{\bar{u}+\bar{d}}{2}+c+t,\\
q^0 &=& \frac{u+d}{2}(L_u^2+L_d^2)+\frac{\bar{u}+\bar{d}}{2}(R_u^2+R_d^2)\nonumber\\
&& +(s+b)(L_d^2+R_d^2)+(c+t)(L_u^2+R_u^2),\\ 
\bar{q}^0 &=& \frac{u+d}{2}(R_u^2+R_d^2)+\frac{\bar{u}+\bar{d}}{2}(L_u^2+L_d^2)\nonumber\\
&& +(s+b)(L_d^2+R_d^2)+(c+t)(L_u^2+R_u^2), 
\end{eqnarray}
with $L_u=1-(4/3)x_W,~L_d=-1+(2/3)x_W,~R_u=-(4/3)x_W$ and $R_d=(2/3)x_W$ where $x_W=\sin^2\theta_W$, and 
$\theta_W$ is the weak mixing angle. 
For the $\bar \nu N$ cross sections, Eqs.~(\ref{cc}) and (\ref{nc}) are the same but with 
each quark distribution function replaced by the corresponding anti-quark 
distribution function, and vice-versa, i.e.,  
$q\leftrightarrow \bar{q},~q^0\leftrightarrow \bar{q}^0$.  

The differential cross sections for the dominant neutrino-electron scattering processes are given by~\cite{Mikaelian:1980vd} 
\begin{eqnarray}   
\frac{d\sigma_{\bar\nu_e e\to \bar\nu_e e}}{dy} &=& \frac{G_F^2 m_e E_\nu}{2\pi}
\left[\frac{R_e^2+L_e^2(1-y)^2}{\left(1+2m_eE_\nu y/M_Z^2\right)^2}+4(1-y)^2\right.\nonumber\\
&&\left.
\times \frac{1+\frac{L_e\left(1-2m_eE_\nu/M_W^2\right)}{1+2m_eE_\nu y/M_Z^2}}{\left(1-2m_eE_\nu/M_W^2\right)^2+\Gamma_W^2/M_W^2}
\right],\label{eq:nue1} \\
\frac{d\sigma_{\bar\nu_e e\to \bar\nu_\mu \mu}}{dy} &=& \frac{G_F^2 m_e E_\nu}{2\pi}\frac{4\left(1-m_\mu^2/2m_eE_\nu\right)^2}{\left(1-2m_eE_\nu/M_W^2\right)^2+\Gamma_W^2/M_W^2},\label{eq:nue2} \\
\frac{d\sigma_{\bar\nu_e e\to \bar\nu_\tau \tau}}{dy} &=& \frac{G_F^2 m_e E_\nu}{2\pi}\frac{4\left(1-m_\tau^2/2m_eE_\nu\right)^2}{\left(1-2m_eE_\nu/M_W^2\right)^2+\Gamma_W^2/M_W^2},\label{eq:nue3} \\
\frac{d\sigma_{\bar\nu_e e\to {\rm had}}}{dy} &=& \frac{d\sigma_{\bar\nu_e e\to \bar\nu_\mu \mu}}{dy}\frac{\Gamma(W\to {\rm hadrons})}{\Gamma(W\to \mu\bar{\nu}_\mu)},
\label{eq:nue4}
\end{eqnarray}
where $L_e=2x_W-1$ and $R_e=2x_W$ are the chiral couplings of $Z$ to electron, and $\Gamma_W$ GeV 
is the total width of the $W$-boson. 

\begin{figure*}[t]
\centering
\includegraphics[width=7cm]{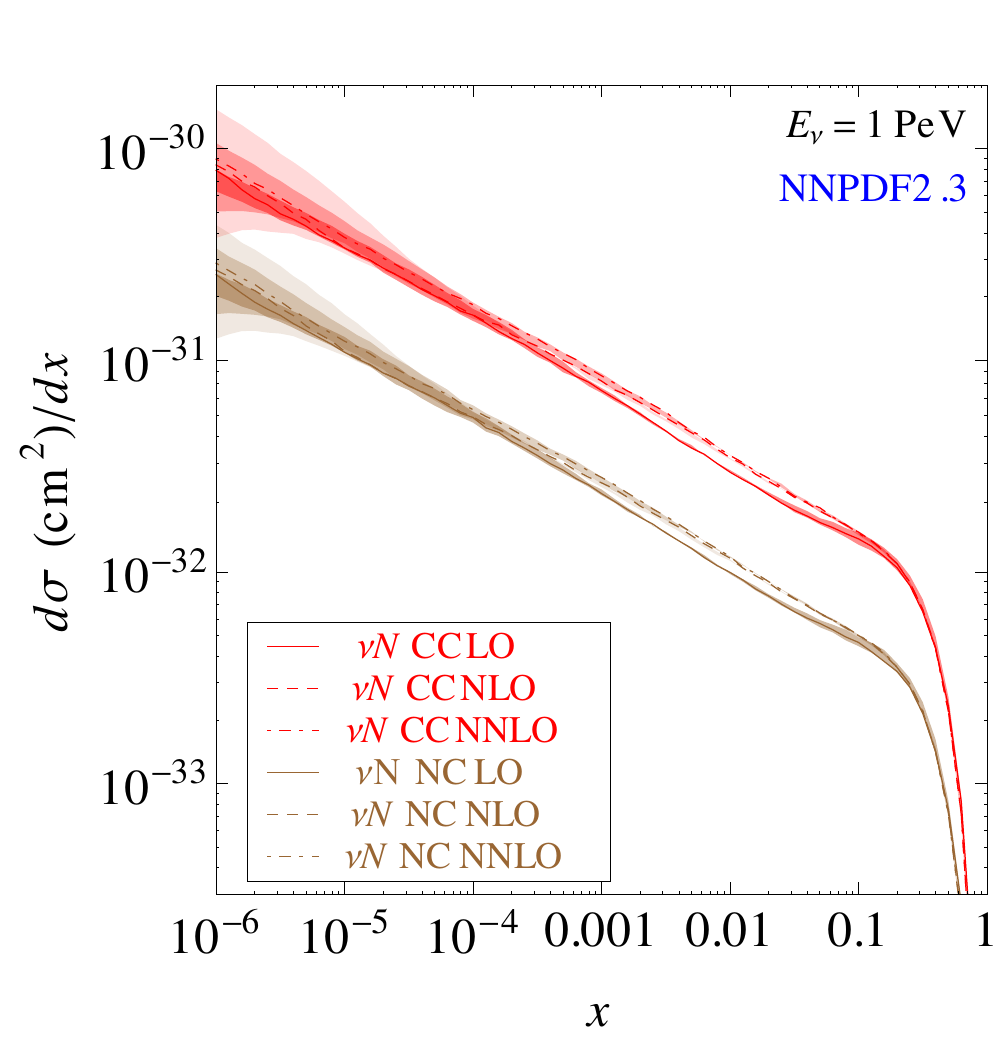}
\hspace{0.5cm}
\includegraphics[width=7cm]{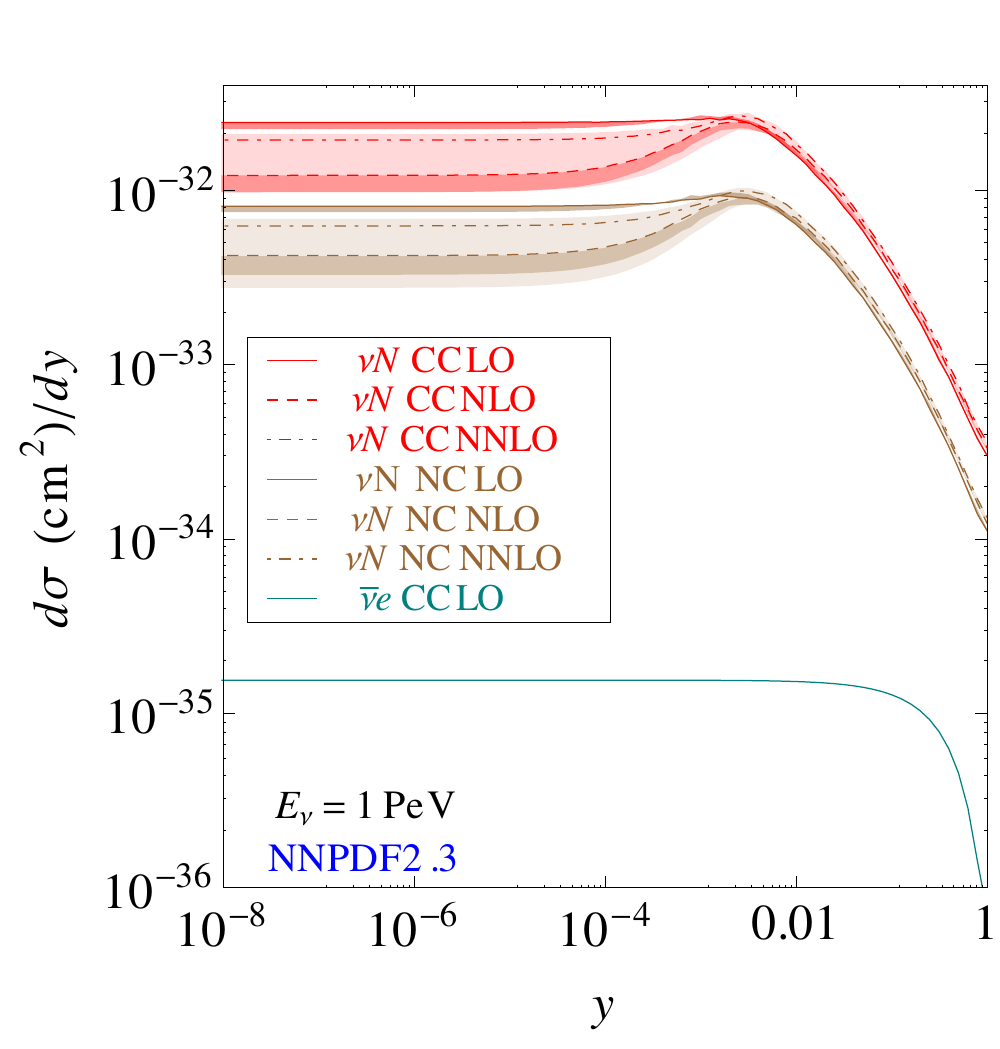}
\caption{The $x$- and $y$-distributions for the neutrino-nucleon CC and NC cross sections for a fixed energy $E_\nu=1$ PeV. The solid, dashed and dot-dashed lines correspond to the distributions with the PDFs at LO, NLO and NNLO respectively, and their 90\% CL uncertainties are shown by the dark to light shaded regions. The $y$-distribution of the electron antineutrino-electron cross section is also shown (lower solid line on the right panel). }
\label{fig1a}
\end{figure*}

The main source of theoretical uncertainties in the neutrino-nucleon 
differential cross sections given by Eqs.~(\ref{cc}) and (\ref{nc}) comes from the PDFs. The size of the PDF uncertainty with respect to the $x$ and $y$ variables defined in Eq.~(\ref{xy}) can be seen from the 
distributions given in Fig.~\ref{fig1a}. For concreteness, we have shown the results for a fixed incoming neutrino energy $E_\nu=1$ PeV, and have used the NNPDF2.3 PDF sets~\cite{Ball:2012cx} based on a 
global data set including the recent LHC data. We use the PDF sets (central values and 90\% CL error) given for $\alpha_s(M_Z)=0.118$, and compare the cross section results evaluated with these PDFs at LO, NLO and NNLO, as shown in Fig.~\ref{fig1a} by the solid, dashed and dot-dashed lines respectively, along with their corresponding 90\% CL bands shown by the (dark to light) shaded regions. We also show the $y$-distribution of the electron antineutrino-electron cross section ($\bar{\nu}_ee\to$ anything) given by Eqs.~(\ref{eq:nue1})-(\ref{eq:nue4}), which is of course independent of the PDF uncertainties. The distributions for the antineutrino-nucleon cross sections are similar to those for the  neutrino-nucleon cross sections, just with slightly smaller values, and are not shown here.    

From Fig.~\ref{fig1a}, we find that the PDF uncertainties in the $y$-distributions are constant in the 
low-$y$ region, while they grow for smaller values of $x$. This is due to 
the uncertainties in the shape of light-quark and gluon distributions in the small-$x$  and high $Q^2$ 
region. The lowest $x$ and highest $Q^2$ scales accessed to date are by the DIS fixed target experiments at 
HERA~\cite{hera}. Including these DIS data in their global PDF analysis, NNPDF2.3 could go down to $x_{\rm min}=10^{-9}$ in the $x$-grid, and up to $Q^2_{\rm max}=10^8~{\rm GeV}^2$ in the $Q$-grid~\cite{Ball:2012cx}. The cross sections calculated using these PDFs have significantly reduced errors at low $x$, as compared to previous analyses (see e.g.,~\cite{gandhi, conolly}).  Similar improved results in the low-$x$ regime were obtained in Ref.~\cite{CooperSarkar:2011pa} using the HERAPDF1.5~\cite{hera, CooperSarkar:2010wm} PDF sets. 
\subsection{Total Cross Section}\label{sec:2b}
\begin{figure}[t]
\centering
\includegraphics[width=7cm]{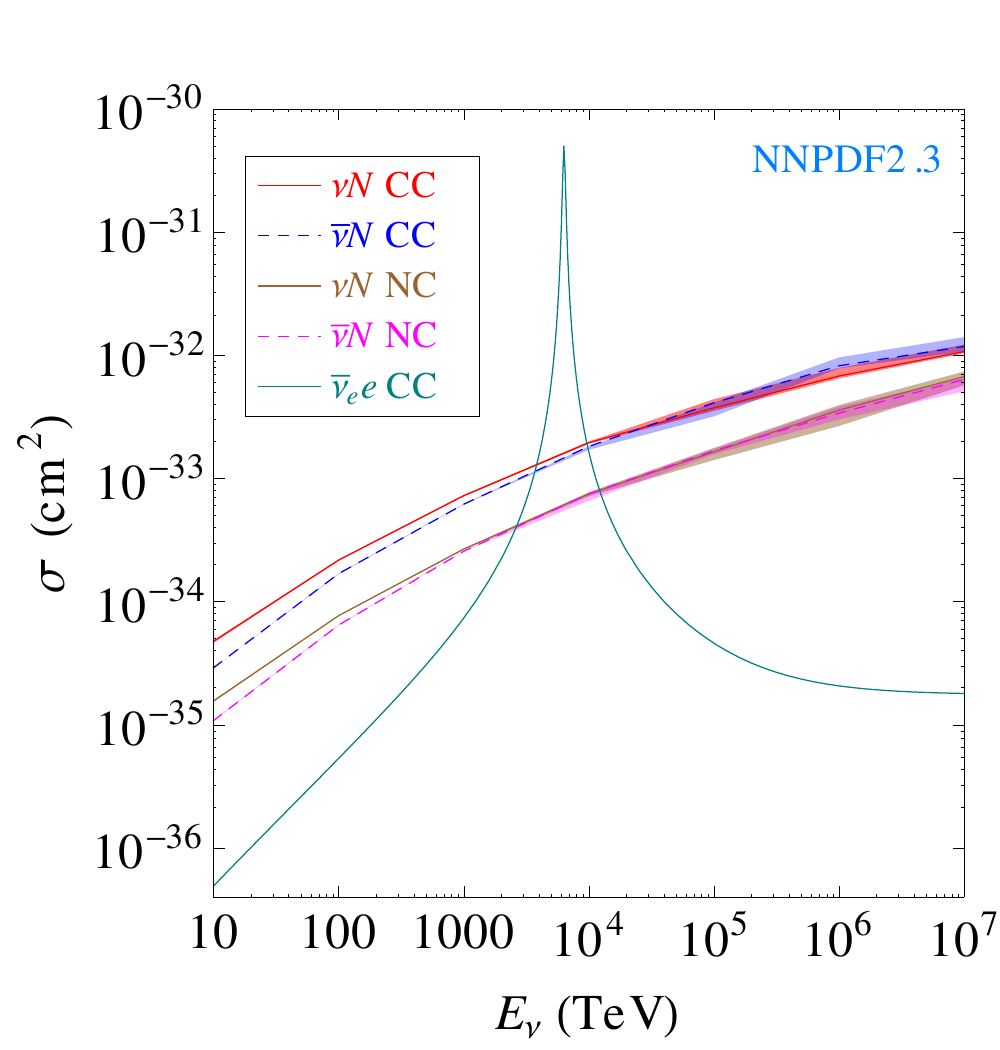}
\caption{The neutrino-induced scattering 
cross sections for the dominant SM processes as a function of the {\it incoming} neutrino energy. 
The $\nu N$ and $\bar\nu N$ cross sections were computed using NNPDF2.3 at NNLO, and their 90\% CL ranges are shown by the shaded regions. }
\label{fig1}
\end{figure}

The total neutrino-nucleon cross section is obtained by integrating the differential cross sections in Eqs.~(\ref{cc}) and (\ref{nc}) over the $x$ and $y$ variables:
\begin{eqnarray}
\sigma(E_\nu) \equiv \int_0^1\int_0^1 dxdy\frac{d^2\sigma}{dxdy},
\label{eq:sigtot}
\end{eqnarray} 
For completeness, we show in Fig.~\ref{fig1} the integrated SM cross sections for the CC and NC 
neutrino-nucleon ($\nu N$) and antineutrino-nucleon ($\bar \nu N$) interactions, computed using the NNPDF2.3 PDF sets at NNLO. The numerical integrations in Eq.~(\ref{eq:sigtot}) were carried out using an adaptive 
Monte Carlo routine. For our numerical purposes, we take the lower limit of the $x$-integration 
to be $10^{-6}$ in order to avoid large uncertainties in the low-$x$ grids.  The shaded regions in Fig.~\ref{fig1} reflect the 90\% CL PDF uncertainties in the total cross section, and also to some extent, the uncertainties from the precision of the numerical integration technique used. Our results for the total cross section agree well with those calculated using other PDF sets; for a comparison, see e.g.,~\cite{conolly, CooperSarkar:2011pa}. 

At very high neutrino energies, the cross sections are dominated by contributions from very small $x$, which currently have a large uncertainty directly associated with the underlying QCD dynamics at high energies~\cite{Formaggio:2013kya}. In this regime, one might have to go beyond the DGLAP 
formalism~\cite{dglap} used by conventional PDF fits, and to consider the 
non-linear evolution of the parton distribution 
arising due to the physical process of 
recombination of partons in the parton cascade. This can be characterized by the saturation scale of the growth of the parton distribution, forming a Color Glass Condensate~\cite{Gelis:2010nm}. Such saturation effects lead to slightly higher values of the neutrino-nucleon cross section and a 
steeper energy dependence at very high energies 
($E_\nu\gsim 100$ PeV)~\cite{Goncalves:2010ay}. 
However, since the current IceCube events are observed at PeV scale and below, 
these non-linear effects are of less importance, and hence, we do not include them in our analysis. 

At higher orders in QCD, the expressions (\ref{cc}) and (\ref{nc}) must be convoluted with appropriate quark and anti-quark density distributions. The heavy quark masses should be taken into account at higher energies in the calculation of the structure functions~\cite{Jeong:2010za}, but the LO cross sections still give us a good estimate of the dominant contributions up to PeV energies. In fact, the numerical values of the cross sections at LO as shown in Fig.~\ref{fig1} agree with the NLO results given in Ref.~\cite{CooperSarkar:2011pa} up to 5\% or so for the current IceCube energy range of interest.  

In Fig.~\ref{fig1} we also show the total $\bar\nu_ee^-$ scattering cross 
section which has the `Glashow resonance' at 6.3 PeV, and gives the dominant contribution in the vicinity of the resonance energy. Other neutrino-electron cross sections are 
not shown here since they are many orders of magnitude smaller than the $\nu N$ cross sections~\cite{Formaggio:2013kya}. The `Glashow resonance' option has been considered earlier~\cite{Bhattacharya:2012fh} 
to explain the two PeV events at IceCube. However, this possibility was disfavored by a dedicated follow-up analysis~\cite{Aartsen:2013bka}. The average incoming neutrino energies required to explain the two PeV events are found to lie below the Glashow resonance region, and hence, the contributions from 
$\bar\nu_e e^-$ scattering to the total number of events turn out to be 
negligible (see Section~\ref{sec:3}). This effect could however be 
important if an excess of events is observed in the 3 - 6 PeV deposited 
energy range.  
\section{SM Prediction for the Number of Events} \label{sec:3}
To the best of our knowledge, no attempt has been made so far to quantify the 
PDF uncertainty effects on the number of events expected in {\it each} of the 
deposited energy bins at IceCube. It is important to include these effects for 
a better  comparison of the observed IceCube events with any particle physics 
explanation. In order to obtain a reliable estimate of the number of IceCube 
events expected due to the SM interactions, we determine the values of various parameters in Eq.~(\ref{eq:N}) as follows: 
\begin{itemize} 
\item $T$=662 days for the IceCube data collected between May 2010 and May 2012~\cite{science}. 
\item $N_{\rm eff}(E_\nu)=N_A V_{\rm eff}(E_\nu)$ where $N_A= 6.022\times 10^{23}~{\rm mol}^{-1}$ is the Avogadro number which is equal to $6.022\times 10^{23}~{\rm cm}^{-3}$ water equivalent for interactions with the ice nuclei.  For interactions with electrons, $N_A$ should be replaced with $(10/18)N_A$ for the number of 
electrons in a mole of H$_2$O. Note that a natural ice nucleus with 10 protons and 8 neutrons is close to being isoscalar, and hence, we use the generic reference PDF sets, without treating the protons and neutrons 
separately. 
\item $V_{\rm eff}(E_\nu)=M_{\rm eff}/\rho_{\rm ice}$  is the effective neutrino target volume, where $\rho_{\rm ice}=0.9167~{\rm g}\cdot{\rm cm}^{-3}$ is the density of ice, and $M_{\rm eff}$  is the effective 
target mass which includes the background rejection cuts and event containment criteria. It depends on the incoming neutrino energy and attains its maximum value $M_{\rm eff}^{max}\simeq 400$ Mton (corresponding to $V_{\rm eff}^{\rm max}\simeq 0.44~{\rm km}^3$ water-equivalent) above 100 TeV for $\nu_e$ CC events~\cite{science}, and above 1 PeV for other CC and NC events. There is some flavor bias at low energies caused by the deposited energy threshold due to missing energy in escaping particles from $\nu_\mu$ and $\nu_\tau$ CC events as well as all flavor NC events, which decreases $M_{\rm eff}$ for these events as compared to the $\nu_e$ CC events. 
\item For the incoming neutrino flux, we assume a Fermi-shock astrophysical flux falling as an 
unbroken power-law spectrum: 
\begin{eqnarray}
\Phi(E_\nu) = CE_\nu^{-s}
\label{phi}
\end{eqnarray}
 for the entire energy range of interest. The exact energy dependence governed by the spectral index $s$ largely depends on the 
extra-terrestrial source evolution models. For a given value of $s$,  the flux normalization $C$ should be chosen to be consistent with the observational upper bound on the fluxes. 
Following the previous IceCube analyses~\cite{Aartsen:2013bka, science}, 
we first show our results for $s=2$ with an all-flavor normalization 
\begin{eqnarray}
C=3.6\times 10^{-8}~{\rm GeV}\cdot {\rm sr}^{-1}\cdot{\rm cm}^{-2}\cdot {\rm s}^{-1},
\label{fnorm}
\end{eqnarray} 
which is the integral upper limit on the UHE cosmic neutrino flux obtained in a previous IceCube search~\cite{Abbasi:2011ji}. This normalization includes equal strength of neutrinos and antineutrinos summed over all neutrino flavors, and assumes an equal flavor ratio of $\nu_e:\nu_\mu:\nu_\tau=1:1:1$ (same for antineutrinos), since neutrino oscillations over astronomical distances tend to equalize the neutrino flavors reaching the Earth, regardless of the initial flux composition~\cite{Choubey:2009jq}. We will also perform a $\chi^2$-analysis with the existing IceCube data to find the best-fit value of the flux normalization for different spectral indices. 
\item The solid angle of coverage $\Omega=2\pi~{\rm sr}$ for an isotropic 
neutrino flux in the southern hemisphere (downward events at IceCube), while for those coming from the northern 
hemisphere (upward events) we must take into account the attenuation effects due to scattering within the Earth which can be represented by multiplying $\Omega$ with an 
energy-dependent shadow factor~\cite{gandhi, gandhi2}
\begin{eqnarray}
S(E_\nu) = \int_{-1}^0 d(\cos\theta)~{\rm exp}\left[-\frac{z(\theta)}{L_{\rm int}(E_\nu)}\right], 
\end{eqnarray}
where $\theta$ is the angle of incidence of the incoming neutrinos above nadir, $z(\theta)$ is the effective column depth which represents the amount of material encountered by an upgoing neutrino in its passage through the Earth, and $L_{\rm int}(E_\nu)=1/\sigma N_A$ is the interaction length. The Earth attenuation effects are relevant at energies above 100 TeV. For the upgoing $\bar{\nu}_e$'s, 
the interaction length is very small near the Glashow resonance (due to its enhanced cross section), and hence, these $\bar\nu_e$'s do not survive their passage through the Earth to the detector. 
For the upgoing $\nu_\tau$'s, there is significant energy loss due to  
regeneration effects inside the Earth, which leads to fast $\tau$-decays producing 
secondary neutrinos (of all flavor) with lesser energy than the original incident one~\cite{Beacom:2001xn}, thereby shifting the energy of the upgoing $\nu_\tau$'s downward when they reach the detector.  
\item The visible energy relevant for detection is the EM-equivalent deposited energy $E_{\rm EM}$ in Eq.~(\ref{eq:N}), which is {\it always} smaller than the incoming neutrino energy $E_\nu$ by a factor which 
depends on the interaction channel. For NC events given by Eq.~(\ref{nc1}), the cross section is identical 
for all flavors, and the fraction of energy imparted to the outgoing hadrons 
$X$ is determined by the inelasticity parameter $y$.  
The resulting hadronic shower yields fewer number of photo-electrons than an equivalent-energy 
electromagnetic shower by a factor $F_X$~\cite{Kowalski:2004qc} which is a function of the hadronic final state energy $E_X=y E_\nu$. We parametrize this energy dependence as~\cite{Gabriel:1993ai} 
\begin{eqnarray}
F_X=1-\left(\frac{E_X}{E_0}\right)^{-m}(1-f_0),
\end{eqnarray}
where 
$E_0=0.399$ GeV, $m=0.130$ and $f_0=0.467$ are the best-fit values from the simulations of a hadronic vertex cascade~\cite{Kowalski:2004qc}. 

Thus for NC events, the total deposited EM-equivalent energy is given by 
$E_{\rm EM, had} = F_X y E_\nu$. 
On the other hand, for $\nu_e N$ CC events given by Eq.~(\ref{cc}), the final state 
electron deposits its entire energy, $E_{\rm EM, e}=(1-y)E_\nu$ into an electromagnetic shower, and there is also an accompanying hadronic shower with deposited energy $E_{\rm EM, had}$. The factor $F_X$ reduces the deposited energy for a hadronic shower to about 80 - 90\% of an equivalent-energy EM shower. 
 
The $\nu_\mu N$ CC events are similar in properties to those due to $\nu_e N$ CC, assuming the final state muon energy to be completely measurable. We have not included in our analysis the effects of muon energy loss during its propagation in rock outside the detector since the IceCube search only considered the interaction vertices well contained within the detector volume, and discarded the events with through-going muon tracks originated outside the detector in order to remove the cosmic ray muon background. 

The $\nu_\tau N$ CC events are however more complicated, with 
properties somewhat between NC and $\nu_e N$ CC events. At the relevant energies (50 TeV$\lsim E_\nu\lsim 2$ PeV), tau leptons will travel only about 10 - 50 m before decaying, so we do not expect them to produce the characteristic ``double bang" signature~\cite{Learned:1994wg} at IceCube as it has a string separation of 125 m~\cite{design}. These distinct signatures may only be visible at IceCube for $\tau$-energies exceeding about 5 PeV when they travel far enough that the cascades from their production and decay are well-separated. The ``double bang" could still be observed with less energetic $\tau$'s 
in the proposed DeepCore experiment with string spacings as small as 42 m~\cite{deepcore}. About 80\% of the $\tau$-decays in the current sample result in a shower, with decays to electrons in an EM shower, and hadronic decays involving multiple mesons in a hadronic shower. The rest $\sim 20\%$ of the taus will produce muons which will give distinct muon tracks. The hadronic showers due to $\tau$-decay will deposit an energy of roughly $(1/2)F_X(1-y) E_\nu$ (the other half being carried away by the associated $\nu_\tau$'s), whereas the leptonic final states will deposit roughly $(1/3)(1-y)E_\nu$,  
the rest being carried away by the final-state neutrinos. 
\end{itemize}
\begin{figure*}[t]
\centering
\includegraphics[width=5.9cm]{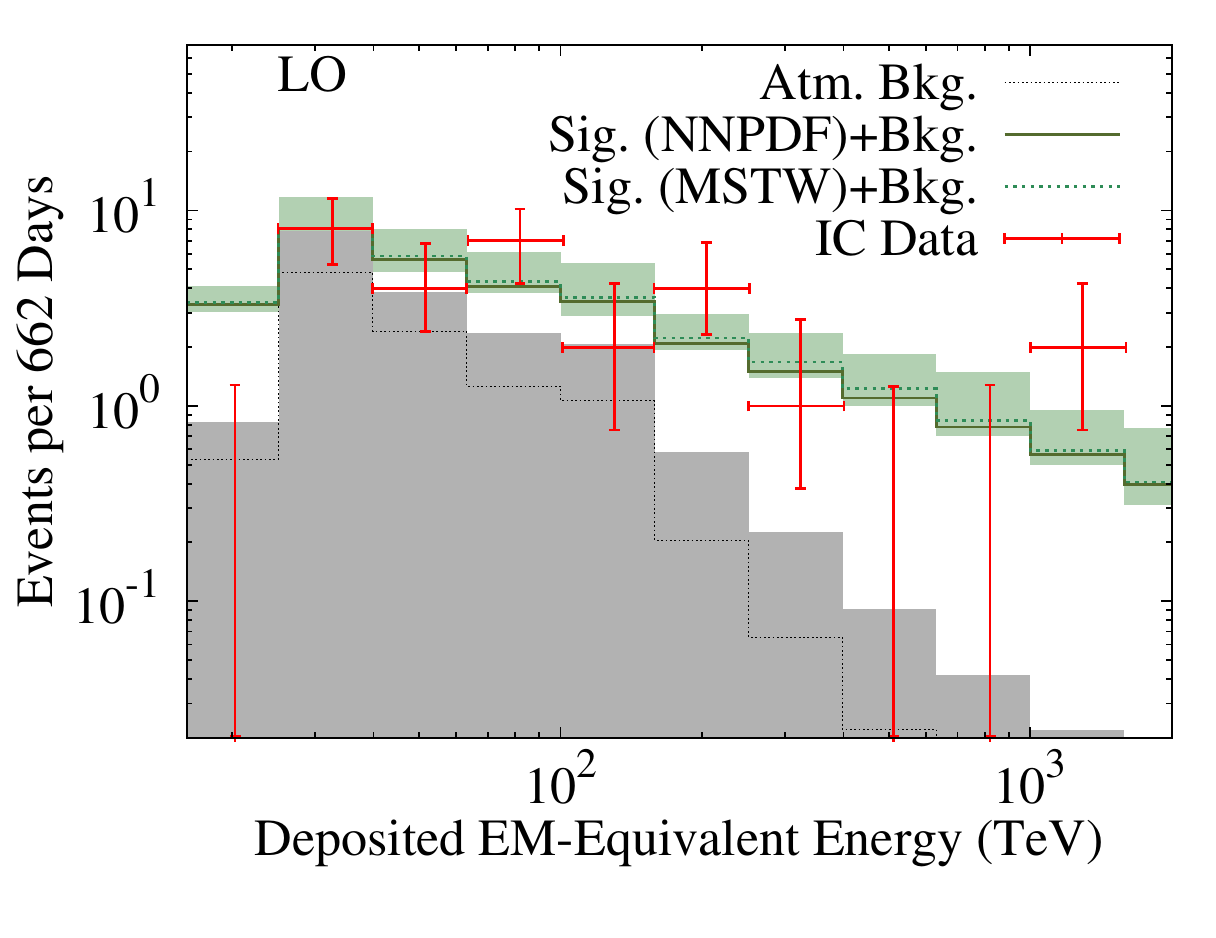}
\includegraphics[width=5.9cm]{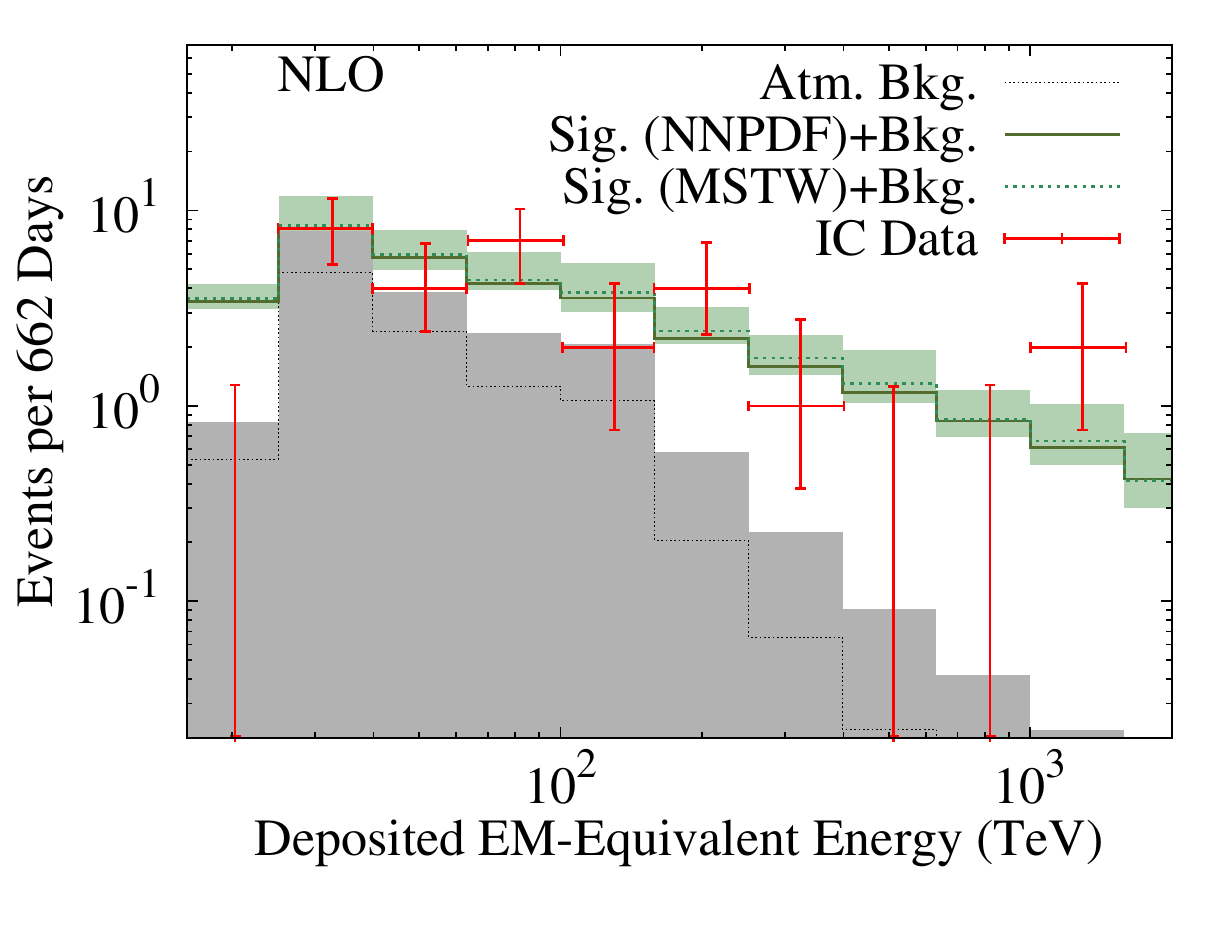}
\includegraphics[width=5.9cm]{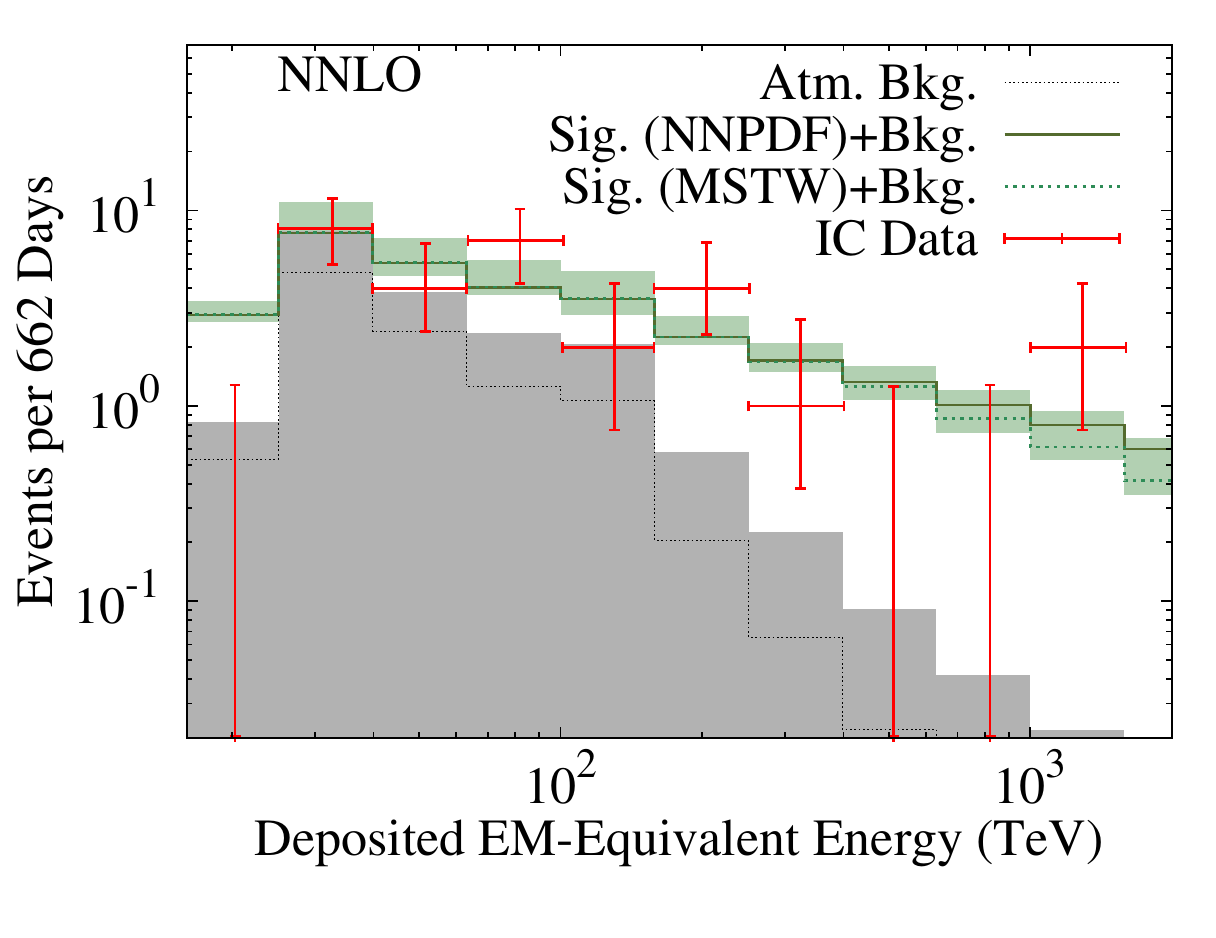}
\caption{The SM signal+background events, along with their 90\% CL PDF 
uncertainties (green shaded), for the IceCube deposited energy 
bins between 16 TeV - 2 PeV. The IceCube data points (with error bars) and the 
atmospheric background 
(black shaded) 
were taken from Ref.~\cite{science}. 
The SM signal events were computed for an $E^{-2}$ flux using the procedure outlined in 
the text, and using two different PDF sets (NNPDF2.3 and MSTW2008) at LO, NLO and NNLO 
for a better comparison. }
\label{fig2}
\end{figure*} 

Our final results for the expected number of events due to the SM interactions  
in all the 11 energy bins analyzed at IceCube are shown in Fig.~\ref{fig2} for an $E^{-2}$ flux. The expected background due to atmospheric neutrinos and atmospheric muons were taken from Ref.~\cite{science}, along with the combined statistical and systematic uncertainties (black shaded) which total to $10.6^{+5.0}_{-3.6}$ events. The SM predictions for the number of signal events were 
obtained using the NNPDF2.3~\cite{Ball:2012cx} PDF sets, and to check the 
robustness of these results, we compared the results computed using the MSTW2008 PDFs~\cite{Martin:2009iq}. We also compare the results obtained with the two PDF sets at LO, NLO and NNLO, including in our analysis 
the 90\% CL PDF uncertainties on the predicted signal+background events for each bin. We find that our signal+background fit is consistent with the IceCube observed data points within their current uncertainties for all the energy bins, except the very first one where we predict an excess of about 2 events over that observed. This is slightly different from the 
IceCube signal+background best-fit given in Ref.~\cite{science}. This mild disagreement may be due to one or several of the following reasons: (i) The IceCube best-fit was derived from a global fit of the deposited energy and zenith distribution of the data to a combination of the atmospheric neutrino background and an isotropic astrophysical flux in the range of 60 TeV - 2 PeV, which does not include the lowest energy bin shown in Fig.~\ref{fig2}. (ii) The PDF uncertainties in the cross section which are not shown in the IceCube best-fit.  (iii) The uncertainty in the flux normalization. The IceCube fit has taken the per-flavor normalization to be $E_\nu^2\Phi(E_\nu)=(1.2\pm 0.4)\times 10^{-8}~{\rm cm}^2\cdot{\rm s}^{-1}\cdot{\rm sr}^{-1}$, whereas our results are derived assuming the central value of $1.2\times 10^{-8}~{\rm cm}^2\cdot{\rm s}^{-1}\cdot{\rm sr}^{-1}$ which is the current upper limit for an equal-flavor composition. (iv) some additional experimental effects relevant at 
lower energies to reduce the atmospheric background 
(e.g., hit probability) not captured in our simple analysis.  It is also 
important to note here that we have directly used the true value of the 
inelasticity parameter $y$ for a given PDF set in our numerical analysis, and not the the average inelasticity parameter as used in some of the previous analyses.


The total number of SM signal events in each channel over the entire energy range of interest shown in Fig.~\ref{fig2} are summarized in Table~\ref{tab1}. The central values and the theoretical errors shown here are derived using the NNPDF2.3 NNLO and its 90 \% CL uncertainties. The corresponding numbers for LO and NLO and also for MSTW PDFs are of similar magnitude and are not shown here.  
Note that at the moment, the IceCube detector can not distinguish between electromagnetic and 
hadronic shower events, and hence, collectively calls them the `cascade' 
events, whereas the muons appear as distinct `track' events. Thus we find that in the energy range of interest, the SM predicts $15.76_{-0.66}^{+1.78}$ cascade events and $5.09_{-0.59}^{+0.36}$ muon tracks. Combining this with the 10.6$^{+5.0}_{-3.6}$ background events, we obtain a total of $31.46_{-4.85}^{+7.13}$ signal+background events, which is consistent with the  28 events observed by IceCube. From the energy distribution shown in Fig.~\ref{fig2}, we find that although the SM expectations for the number of events in the highest energy bin observed so far is slightly below the current experimental central value, it is still consistent within the theoretical and experimental uncertainties. Note however that the SM fit shown in Fig.~\ref{fig2} was obtained with the current {\it upper} limit on the normalization of an astrophysical $E^{-2}$ flux; 
hence, any additional excess in the future data and/or improvement in the flux limit would make it 
extremely difficult to explain by the SM alone, and could give us an important clue to some new physics.   
\begin{table}[t]
\begin{center}
\begin{tabular}{c|c|c|c|c}\hline\hline
channel & hadron & electron & muon & total \\ \hline
$(\nu+\bar\nu) N $ NC & $1.54_{-0.14}^{+0.12}$ & - & - &  $1.54_{-0.14}^{+0.12}$ \\
$(\nu_e+\bar\nu_e) N$ CC & $2.42_{-0.09}^{+0.30}$ & $6.74_{-0.13}^{+0.75}$ & - & $9.15_{-0.22}^{+1.05}$\\
$(\nu_\mu+\bar \nu_\mu) N$ CC & $1.62_{-0.06}^{+0.22}$ & - & $4.39_{-0.12}^{+0.53}$ & $6.01_{-0.18}^{+0.75}$ \\
$(\nu_\tau+\bar\nu_\tau) N$ CC & $3.05_{-0.07}^{+0.40}$ & $0.23_{-0.00}^{+0.03}$ & $0.22_{-0.00}^{+0.03}$ & $3.51_{-0.08}^{+0.47}$ \\
$\bar{\nu}_e e$ & 0.09 & 0.01 & 0.01 & 0.11\\ \hline 
total SM & $8.78_{-0.51}^{+1.00}$ & $6.99_{-0.15}^{+0.77}$ & $5.09_{-0.59}^{+0.36}$  & $20.86_{-1.25}^{+2.13}$ \\ \hline
\hline 
\end{tabular}
\end{center}
\caption{Total number of SM signal events expected from different final states in the deposited energy range 16 TeV - 2 PeV. 
The theoretical errors are derived using the 90\% CL  PDF uncertainties.}
\label{tab1}
\end{table}
\begin{figure*}[t]
\centering
\includegraphics[width=5.9cm]{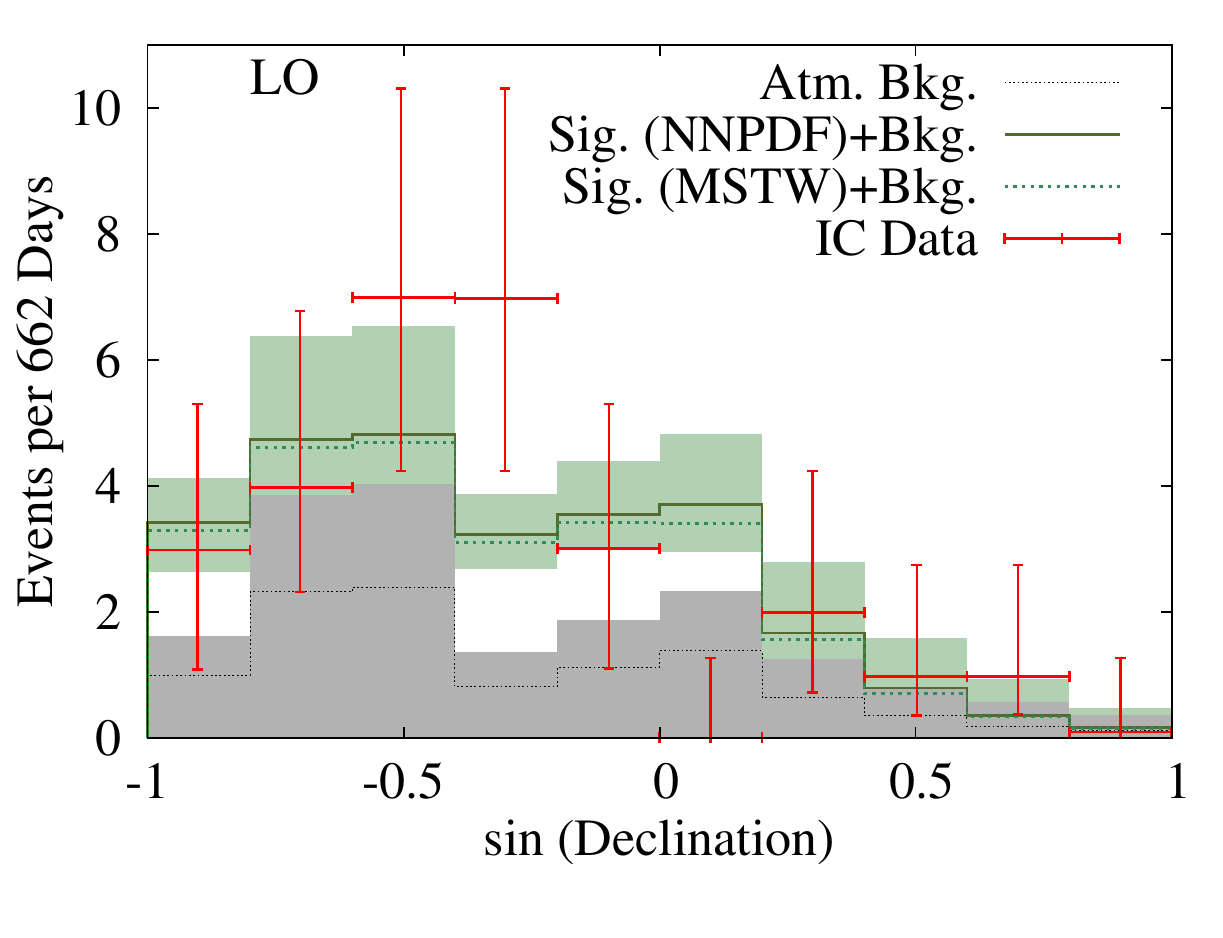}
\includegraphics[width=5.9cm]{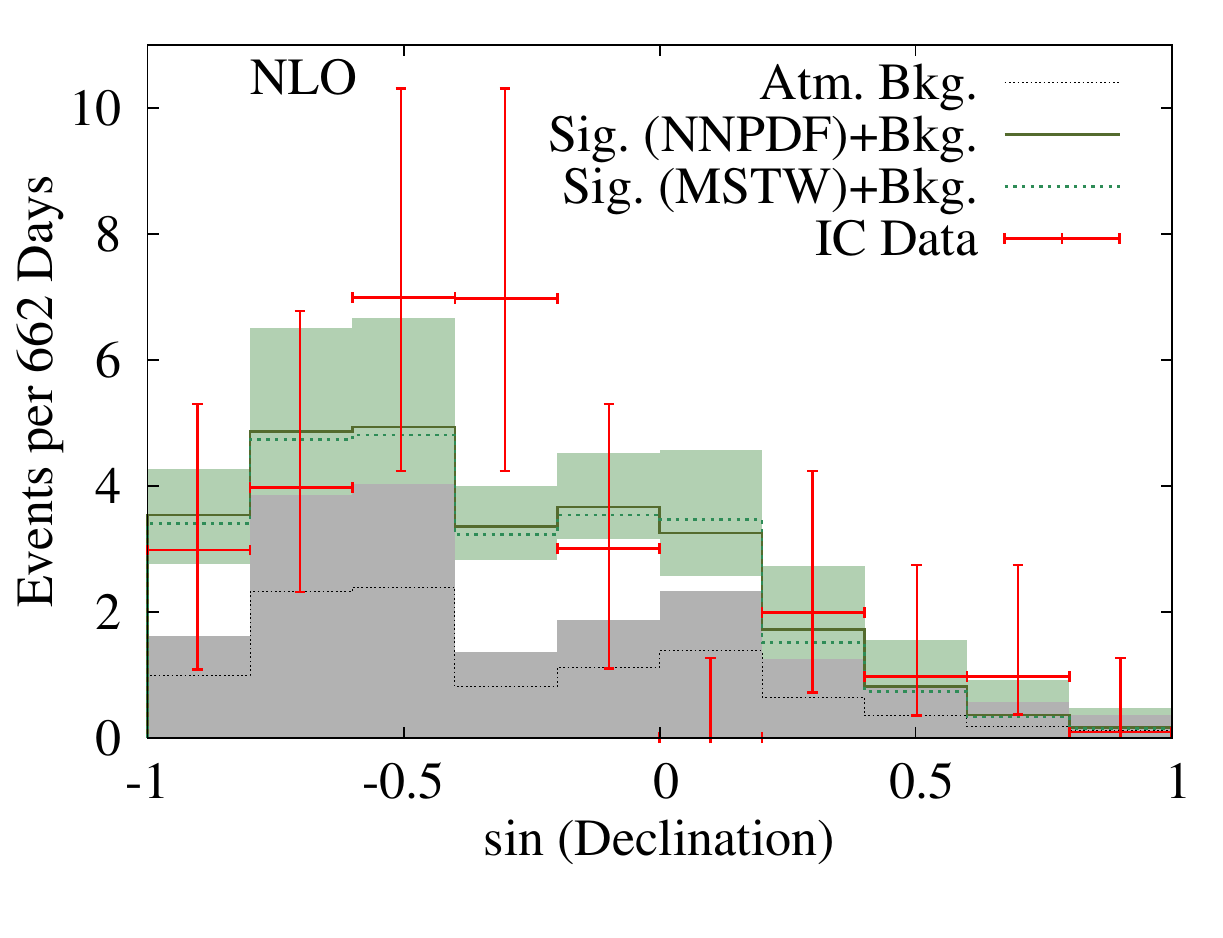}
\includegraphics[width=5.9cm]{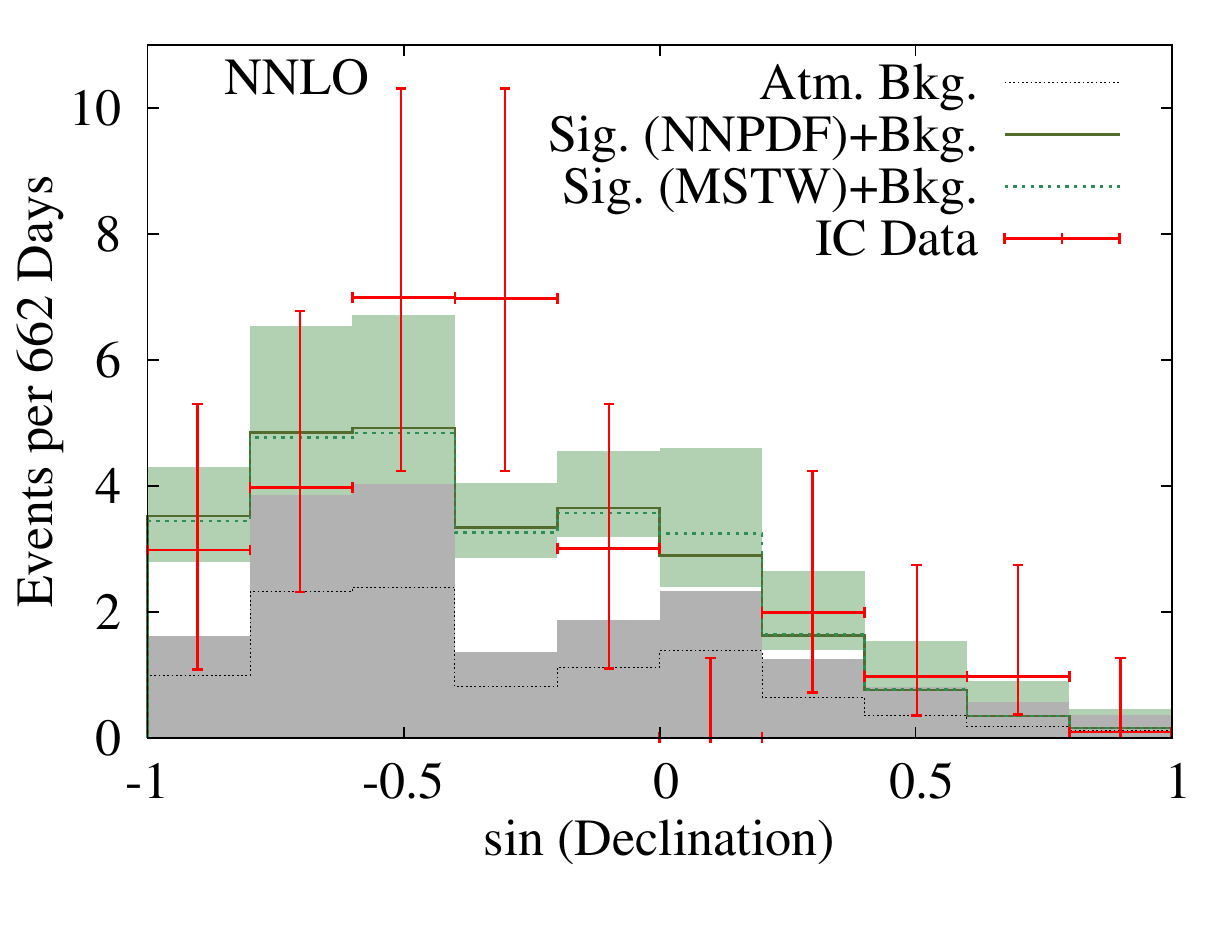}
\caption{The zenith angle distribution of the SM signal+background events, along with the 90\% CL PDF uncertainties (green shaded). The IceCube data points (with error bars) and the atmospheric background 
(black shaded) were taken from Ref.~\cite{science}.}
\label{fig3}
\end{figure*} 
 
In Fig.~\ref{fig3}, we show the distribution of the declination angles of the 
events shown in Fig.~\ref{fig2}, and find that out of the $19.58_{-0.61}^{+1.77}$ signal events, $12.64_{-0.29}^{+0.26}$ are downgoing and $8.21_{-0.96}^{+1.87}$ upgoing.  Combining this information with the distribution of the atmospheric background events, our signal+background fit seems to be in good agreement with the IceCube data obtained so far.  Apart from the deposited energy and declination angle distributions, we can also understand several other features of the IceCube data with our simple SM interpretation: 
\begin{itemize}
\item There are more downgoing events (about 60\%) than upgoing due to the Earth attenuation effects. 
\item The number of muon tracks ($2.89_{-0.06}^{+0.04}$ downgoing and $2.20_{-0.53}^{+0.31}$ upgoing) 
predicted in the SM seems to be consistent with the 7 track-like events (1 upgoing, 6 downgoing) 
observed by IceCube, with 4 of the downgoing events in the lower energy bins 
consistent with the expected $6.0\pm 3.4$ background atmospheric muons. 
So at the moment, there does not seem to be a statistically 
significant paucity of muons. We must however emphasize that if there is indeed a persistent 
paucity of muons 
in future as more data is accumulated, and results in a  significant statistical discrepancy with the SM expectations shown here, one will have to seriously consider some beyond SM explanation. 
For instance, one possible solution in such a (currently hypothetical) scenario could be due to a 
lepton-flavor 
violating (LFV) gauge interaction in a  warped extra-dimensional setup~\cite{Agashe:2006iy}. In these models, one may have $\nu_\mu N\to \nu_\tau X$ CC interactions mediated by a TeV-scale $W'$ in the $t$-channel which could cause the paucity of muon events, while being consistent with the current experimental limits on LFV processes. Another alternative solution to the `muon problem' could be due to 
the presence of an $R$-parity violating interaction in a supersymmetric scenario. 
However, as we have emphasized earlier, it is premature to consider such exotic possibilities without a clear statistical deviations of the IceCube data from the SM expectations.   
\item For the duration of the current data-taking by IceCube, the lack of events between 2 - 10 PeV, 
as would be expected by an unbroken $E^{-2}$ flux considered here, indicates that there might be a break or cut-off in the spectrum close to 2 PeV, or the spectrum could be softer (such as $E^{-2.2}$ or $E^{-2.3}$, but with a higher value of the flux normalization). However, it is difficult to explain {\it all} the observed events with a softer spectrum, as we will see below.  
\end{itemize}

\begin{figure}[t]
\centering
\includegraphics[width=8cm]{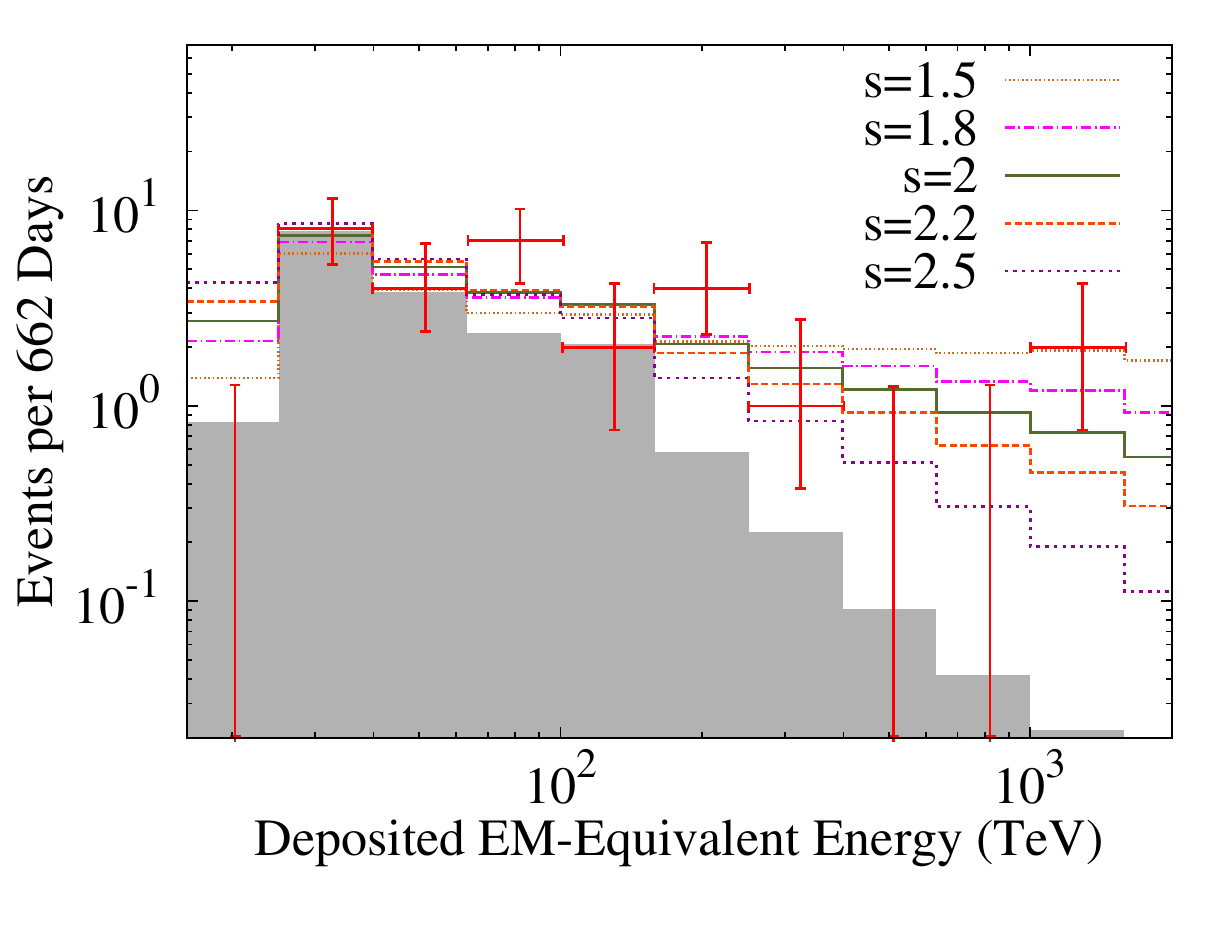}
\caption{The energy distribution of the SM signal+background events (similar to Fig.~\ref{fig2}) 
for a power-law flux with different spectral indices. 
The flux normalization is fixed by minimizing the $\chi^2$-value. For PDFs, we have taken the central values of the NNPDF2.3 NNLO PDF sets. }
\label{fig5}
\end{figure}

In the analysis presented above, we have assumed a simple unbroken power-law flux given by Eq.~(\ref{phi}) with $s=2$ and the flux normalization given by Eq.~(\ref{fnorm}). To ascertain the range of the spectral index $s$ compatible with the existing IceCube data, we perform a $\chi^2$-analysis, with the $\chi^2$-value is defined as 
\begin{eqnarray}
\chi^2=\sum_i\frac{(f_i^{\rm SM}-f_i^{\rm IC})^2}{\delta f_i^2},
\end{eqnarray}
where $f_i^{\rm SM}$ and $f_i^{\rm IC}$ are the number of events in the $i$-th energy bin, as predicted by the SM signal+background and as observed by IceCube respectively, and $\delta f_i$ is the corresponding experimental 
uncertainty in the $i$-th bin as reported by IceCube. The results are 
summarized in Table~\ref{tab2} and also in Fig.~\ref{fig5} for some typical values of $s$. For a given value of $s$, we fix the overall flux normalization $C$ by minimizing the $\chi^2$-value over the 7 energy bins with non-zero observed number of events. The resulting energy distribution is shown in Fig.~\ref{fig5}. Here we have chosen the central values of the NNPDF2.3 NNLO PDF sets. The corresponding PDF uncertainties are similar to those shown in Fig.~\ref{fig2}, and hence, not shown here for purposes of clarity. We find that a softer spectrum ($s>2$) does not give a good fit to the existing data, and the best-fit range of the spectral index is $s=1.8$ - 2, though the current data does not exclude a harder spectrum up to $s=1.5$, provided there is a cut-off in the spectrum close to 2 PeV. The corresponding flux normalization range given in Table~\ref{tab2} is consistent with the current upper bounds from IceCube~\cite{Abbasi:2011ji, Abbasi:2011jx, Abbasi:2012cu}, and could be tested with more data in future.  
\begin{table}[t]
\begin{center}
\begin{tabular}{c|c|c}\hline\hline
$s$ & $\chi^2_{\rm min}$ & $C~({\rm GeV}\cdot {\rm sr}^{-1}\cdot{\rm cm}^{-2}\cdot {\rm s}^{-1})$\\ \hline
1.5 & 0.99 & $1.7\times 10^{-9}$\\
1.8 & 0.87 & $1.0\times 10^{-8}$\\
2.0 & 0.88 & $3.0\times 10^{-8}$\\
2.2 & 0.94 & $8.3\times 10^{-8}$\\
2.5 & 1.11 & $3.2\times 10^{-7}$\\ \hline\hline
\end{tabular}
\end{center}
\caption{The best-fit flux normalization values for different spectral indices of the incoming UHE 
neutrino flux.}
\label{tab2}
\end{table}
 
\section{Conclusion} \label{sec:4}
In summary, for a reliable search for signals of New Physics by the powerful IceCube detector, it is desirable to have a very good understanding of all aspects of the observed UHE neutrino events. Here we have shown that from a particle physics point of view, the current data seems to be consistent with the SM explanation for a simple astrophysical power-law flux, $\Phi=CE_\nu^{-s}$ with $C=$(0.2 - 3)$\times 10^{-8}~{\rm GeV}\cdot {\rm sr}^{-1}\cdot{\rm cm}^{-2}\cdot {\rm s}^{-1}$ and $s=1.5$ - 2, and so far does not require any New Physics. However, it is extremely important to bear in mind that as the statistics solidifies with the accumulation of more IceCube data, it would provide us with a unique test of the SM with the 
highest neutrino energies ever observed in Nature, and if  
 any significant deviation from the current energy spectrum emerges, it will call for a beyond SM scenario. 
 
\section*{Acknowledgments}
We would like to thank Steve Barwick, Francis Halzen, Claudio Kopper, Alexander Mitov, Subir Sarkar, Maria Ubiali and Nathan Whitehorn for very helpful discussions and inputs. 
P.S.B.D. acknowledges the local 
hospitality provided by the High Energy Theory group, Brookhaven National Laboratory,  
where this work was initiated. 
The work of C-Y.C. and A.S. is supported by the US Department of Energy under Grant DE-AC02-98CH10886, and 
P.S.B.D. is supported by the Lancaster-Manchester-Sheffield Consortium for Fundamental Physics under 
STFC grant ST/J000418/1.

\end{document}